\documentclass[utf8]{frontiersSCNS}
\usepackage{url,hyperref,lineno,microtype,subcaption}
\usepackage[onehalfspacing]{setspace}
\usepackage[normalem]{ulem}   

\def\firstAuthorLast{McIntosh {et~al.}} 
\def\Authors{Scott W.\ McIntosh\,$^{1,*}$, Phillip H. Scherrer\,$^{2}$, Leif Svalgaard\,$^{2}$ and Robert J. Leamon\,$^{3,4}$}
\def\Address{
  $^{1}${National Center for Atmospheric Research, P.O. Box 3000, Boulder, CO~80307, USA} \\
  $^{2}${Hansen Experimental Physics Lab, Stanford University, Stanford, CA 94305, USA} \\  
  $^{3}${University of Maryland--Baltimore County, Goddard Planetary Heliophysics Institute, Baltimore, MD 21250, USA} \\
  $^{4}${NASA Goddard Space Flight Center, Code 672, Greenbelt, MD 20771, USA}
  }

\def\corrAuthor{Scott McIntosh}
\def\corrEmail{mscott@ucar.edu}

\newcommand{\degree}{\ensuremath{^\circ}}

\newcommand{\apjl}{   {\it Astrophys. J. Lett.}}

\newcommand{\jgr}{    {\it J. Geophys. Res.}}

\newcommand{\solphys}{{\it Solar Phys.}}

\chardef\us=`\_

\begin{document}
\onecolumn
\firstpage{1}

\title[Uniting Hale and Extended Cycles]{Uniting The Sun's Hale Magnetic Cycle and `Extended Solar Cycle' Paradigms}

\author[\firstAuthorLast ]{\Authors} 
\address{} 
\correspondance{} 

\extraAuth{}

\maketitle

\begin{abstract}
Through meticulous daily observation of the Sun's large-scale magnetic field the Wilcox Solar Observatory (WSO) has catalogued two magnetic (Hale) cycles of solar activity. Those two ($\sim$22-year long) Hale cycles have yielded four ($\sim$11-year long) sunspot cycles (numbers 21 through 24). Recent research has highlighted the persistence of the ``Extended Solar Cycle'' (ESC) and its connection to the fundamental Hale Cycle - albeit through a host of proxies resulting from image analysis of the solar photosphere, chromosphere and corona. This short manuscript presents the correspondence of the ESC, the surface toroidal magnetic field evolution, and the evolution of the Hale Cycle. As Sunspot Cycle 25 begins, interest in observationally mapping the Hale and Extended cycles could not be higher given potential predictive capability that synoptic scale observations can provide.
\end{abstract}

\section*{Introduction}
For over four centuries solar observers have pondered the physical origins of the canonical marker of solar activity - the sunspot. It took more than 200 years after the sketching and cataloging of sunspots commenced before it was discovered that the number of sunspots waxes and wanes over an approximately 11-year period \citep{Schw49}. A half century later, mapping the latitudinal variation of the spotted Sun yielded the ``butterfly diagram,'' a pattern progressing from latitudes around 30$^\circ$ (north and south) to the equator over the ~11-year period \citep{Maun04}. In the golden age of solar astronomy that followed, it was first suggested \citep{1908ApJ....28..315H} and then demonstrated \citep{Hale19} that sunspots were sites of intense magnetism protruding through the Sun's photosphere and that the polarities of the butterfly's wings alternated in sign with a period of about 22 years \citep{1925ApJ....62..270H}. This alternating magnetic polarity cycle is synonymously identified with its discoverer, the eponymous (22-year) ``Hale Cycle,'' or the (22-year) ``Hale Magnetic Polarity Cycle.'' Understanding how the magnetic spots, their butterfly patterning, and the polarity flipping are tied together to drive solar activity has formed the keystone problem of observational \citep{Bab61}, theoretical \citep{Lei69} solar- and astro-physics in the intervening century \citep[e.g.,][]{2010LRSP....7....1H}.

For over four decades another term describing solar activity has sporadically appeared in the literature - the ``Extended Solar Cycle.'' The extended solar cycle \citep[e.g.,][]{1987SoPh..110....1W} (ESC) was used to describe an spatio-temporal extension of the sunspot butterfly pattern to higher solar latitudes (to around 55{$^{\circ}$}) and further back in time (by almost a decade). A culmination of many years of painstaking observation the ESC is exhibited in prominences and filaments  \citep[e.g.,][]{1933MmArc..51....5B,1975SoPh...44..225H}, `ephemeral’ (small-scale transient) active regions \citep[e.g.,][]{1973SoPh...32..389H}, global-scale features of the Sun's corona \citep[e.g.,][]{1988sscd.conf..414A} and the zonal flow patterns \citep[e.g.,][]{1980ApJ...239L..33H,1987Natur.328..696S} of the `torsional oscillation.' In effect, this assortment of observational phenomena created a set of spatio-temporally overlapping chevron-like activity patterns. 

\begin{figure}[ht]
\centering
\includegraphics[width=0.75\linewidth]{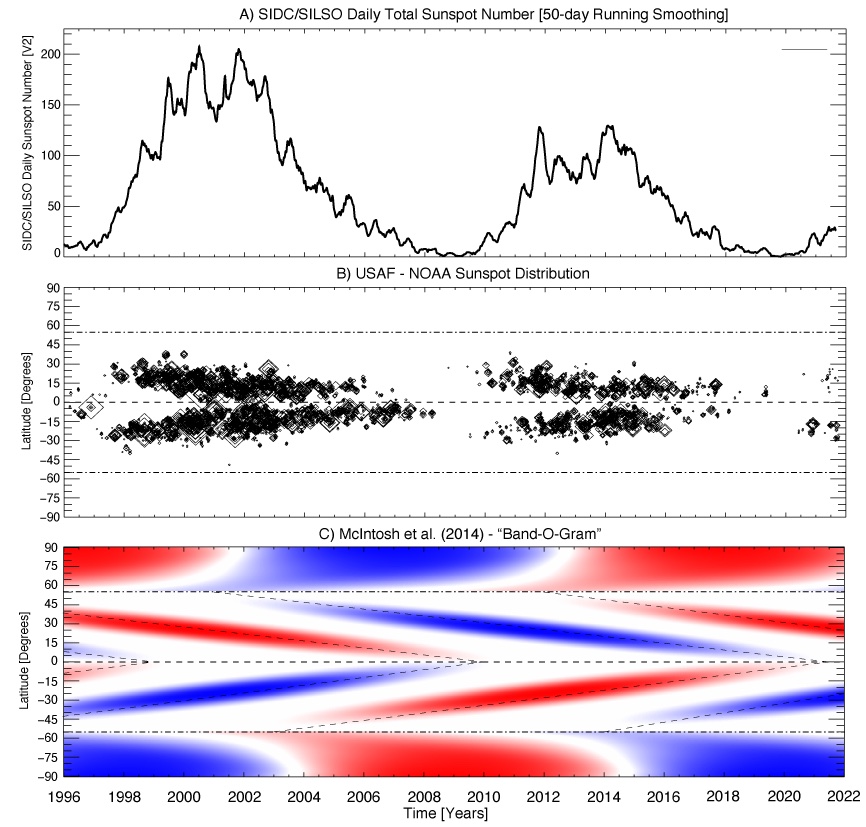}
\caption{Sunspot evolution since 1996. Comparing and contrasting the evolution of the total sunspot number provided (panel A), the spatio-temporal distribution of sunspots provided by the US Air Force and NOAA (panel B), and a data-driven schematic of the Hale Cycle evolution constructed by M2014, the band-o-gram (panel C).}
\label{fig:f0}
\end{figure}

The concept of the ESC was 're-discovered' by McIntosh et al. in their study of extreme ultraviolet brightpoints and their associated magnetic scale \citep[hereafter M2014]{McIntosh2014}. They identified a pattern of coronal and photospheric features that was greatly extended in time and latitude relative to the sunspot butterfly. They deduced that the activity bands observed were the (toroidal) magnetic bands of the Hale Cycle, but no concurrent photospheric magnetic measurement was available to affirm their deduction. The core inference of their study was that the spatio-temporal overlap and interaction of extended activity bands observed contributed directly to the shape (the butterfly) and modulation (the amplitude) of the sunspot cycle. 

Figure~\ref{fig:f0} shows the evolution of the total sunspot number, the latitudinal distribution of sunspots and the data-inspired construct introduced by M2014 that inferred the magnetic activity band arrangement and progression of the Hale Cycle and how those bands contribute to the modulation of sunspot cycles. This `band-o-gram,' introduced in section~3 (and Fig.~8) of M2014, was intended as a qualitative, and not quantitative, illustration of the position, timing and magnetic field strength of the bands---with the emphasis on their phasing. The activity bands in the band-o-gram start their (assumed) linear progression towards the equator from 55\degree latitude at each hemispheric maxima, meeting and disappearing at the equator at the terminator. At the terminator the polar reversal process commences at 55\degree latitude, progressing poleward at their (assumed) linear rate---reaching the pole at the appropriate hemispheric maximum. So, for a list of hemispheric maxima and terminators, a band-o-gram can be constructed. The width of the bands is prescribed by a Gaussian distribution 10 degrees in latitude, commensurate with those observed in the coronal brightpoints originally studied by M2014.

\section*{Data \& Method}
The Wilcox Solar Observatory (WSO) began collecting daily low spatial resolution observations of the Sun's global (or mean) magnetic field in May 1975 \citep{1977SoPh...54..353S} and a very well-known data product of WSO is the evolution of the Sun's polar cap magnetic fields \citep{1978SoPh...58..225S}. These low-resolution synoptic observations are ideal for identifying large-scale, long-lived, patterns - reducing the effects of small-scale, rapidly changing fields of emerging magnetic regions. Following, \cite{1979SoPh...61..233D} the daily WSO magnetograms are obtained by scanning boustrophedonically along 11 east- west rows (i.e., the observation of alternate rows in opposite directions---if one row is taken from left to right then the next row is from right to left). The $180"$ magnetograph aperture moves $90"$ between points in the east-west direction and $180"$ north or south between rows, taking a 15s integration of the Fe~I 5247\AA{} line at 195 points on the solar disk---resulting in a total of about 2 hours per daily map. Because of the large aperture size of the magnetograph the regions from 70\degree{} to the poles lie entirely within the last aperture and are not resolved.

Following the method of \cite{1974SoPh...39..275H} and \cite{1979SoPh...61..233D}, the daily WSO magnetographs can be decomposed into the poloidal and toroidal components which, according to dynamo models, are regenerated from one another, alternating and repeating in an approximately 22-year cycle \citep[e.g.,][]{2010LRSP....7....3C}. The method used to perform this decomposition is detailed by \citep{1994SoPh..153..131S}, where the daily WSO magnetographs are first separated into their positive and negative magnetic field polarities which are then tracked as they cross the solar disk. They are then fitted to estimate the average east-west inclination angle of the magnetic field---or the toroidal component of the photospheric magnetic field \citep[see Fig.~1 of][for an illustration of the geometry]{2010ASPC..428..109L}. 

In this paper we use the \cite{1994SoPh..153..131S} derivative data product of the WSO toroidal magnetic field component in the photosphere and the WSO polar magnetic field estimate using the five central aperture pointings (central meridian $\pm$ two) in first and last rows of observations documented by \cite{1978SoPh...58..225S}.

\begin{figure}[ht]
\centering
\includegraphics[width=\linewidth]{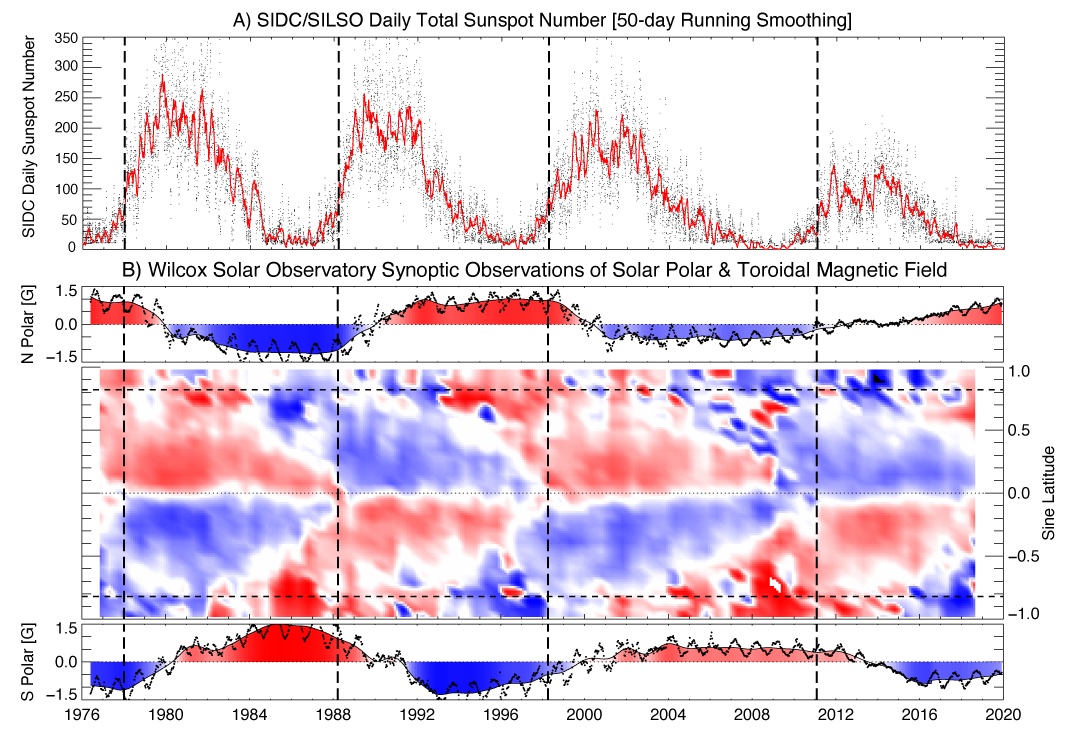}
\caption{WSO Inferred toroidal magnetic field evolution since 1976. Comparing and contrasting the evolution of the total sunspot number (panel A), with the spatio-temporal distribution of the derived toroidal magnetic field component (central) and polar magnetic field components (above north and below south) derived from daily WSO observations (panel B). Note that the toroidal field panel is in its native sine latitude format \citep{2010ASPC..428..109L}. The horizontal dashed lines indicate a latitude of 55\degree{} while the vertical dashed lines shown in each panel mark the times of the Hale Cycle termination events studied by M2019.}
\label{fig:f1}
\end{figure}

\section*{Results}
An initial study of the slowly evolving behavior \citep{1994SoPh..153..131S} noted the potential relationship with the ESC. Figure~\ref{fig:f1} contrasts four and a half decades of WSO observations with the evolution of the sunspot number over the same timeframe. Panel B shows the latitude-time variation of the WSO toroidal magnetic field component in addition to the field strength of the northern and southern polar regions.

Several features of Figure~\ref{fig:f1} are immediately visible, but perhaps the most striking are the strong overlap in time of the toroidal magnetic systems, the short transitions from one polarity to the next \-- evidenced through the narrow white (very near 0G) zones, the lack of field migration across the Sun's equator, and the close association of these last two features at the Sun's equator four times in the record (in 1978, 1988, 1998 and 2011). The patterns, including a strong resemblance to the ESC, are described in more detail by \cite{1994SoPh..153..131S} and \cite{2010ASPC..428..109L}. 

\begin{figure}[ht]
\centering
\includegraphics[width=\linewidth]{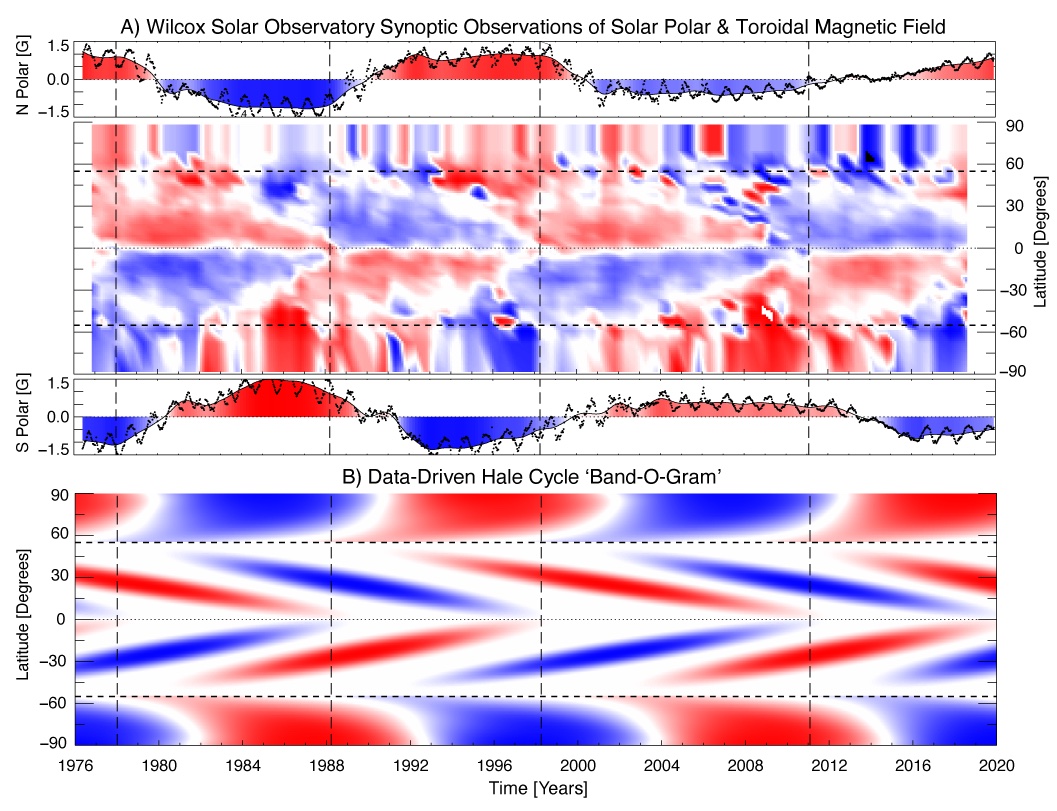}
\caption{Comparing and contrasting the WSO toroidal magnetic field (Panel A, see Figure~\ref{fig:f1}B and now expressed in latitude) and polar cap measurements with the data-inspired band-o-gram (cf. Fig.~\ref{fig:f0}C) now extended to cover the WSO record. The horizontal dashed lines indicate a latitude of 55\degree{} while the vertical dashed lines shown in each panel mark the times of the Hale Cycle termination events studied by M2019.}
\label{fig:f2}
\end{figure}

The last of these features, synchronized zero-crossing transitions at the lowest latitudes in each hemisphere, are concurrent with events that mark the end of the Hale Cycle progressions, or termination events as they have become known, that were initially described by M2014 and explored again (in more detail) recently \citep[][hereafter M2019]{McIntosh2019}. The termination events are illustrated with dashed vertical lines in Figure~\ref{fig:f1}. These events signify the final cancellation of the magnetic systems that were responsible for the last sunspot cycle at the equator and, near-simultaneously, a period of very rapid growth of the next sunspot cycle at mid-solar latitudes. Interestingly, M2019 also noted that these termination events at the equator were co-temporal with the start of the polar magnetic field reversal process. This process is perhaps best visualized through the observed progression of the highest latitude filaments (or polar crown filament) to the pole, the so-called ``rush to the poles'' \citep[e.g.,][]{Bab61, 1989SoPh..119..323S}. The time at which this poleward march completes corresponds to when the measured polar magnetic field crosses zero.

In order to visually compare the WSO observations [Figure~\ref{fig:f1}B] and the ESC band-o-gram [Figure~\ref{fig:f0}C] (extended to cover the baseline of the WSO observations) we convert the WSO data from sine latitude to latitude and the result can be seen in Figure~\ref{fig:f2}.

Additionally, Table~1 of \cite{2022ApJ...927L...2L} places bounds on the correspondence of the toroidal field zero-crossings (near the solar equator) with the Hale Cycle terminator events determined by other means. With the 5 degree resolution of the WSO magnetograph scanning rows around the equator the zero-crossing times of the toroidal magnetic field provided correspond well with the Hale Cycle terminator times presented in Table~1 of M2019: 1978.00 [N5:1976.67, S5:1977.17]; 1988.50 [N5:1988.75, S5:1986.83]; 1997.75 [N5:1997.75, S5:1999.17]; 2011.20 [N5:2008.83, S5:2011.50].

\subsection*{High-Res/Low-Res \&\ The 2021 Hale Cycle Termination}
\begin{figure}[ht]
\centering
\includegraphics[width=\linewidth]{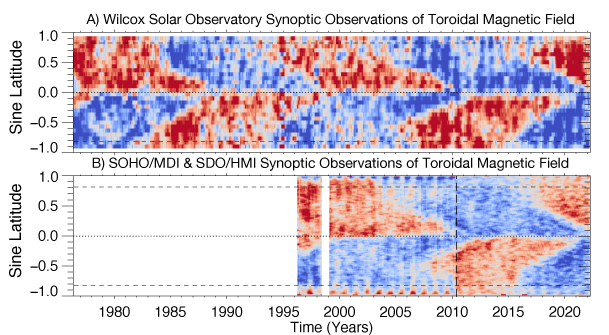}
\caption{Comparing and contrasting the WSO toroidal magnetic field (Panel A) and version derived from the higher-resolution full-disk space observations SOHO/MDI and SDO/HMI (Panel B) with the latter described by \cite{2022ApJ...927L...2L} updated to the present. The vertical dashed line in Panel B indicates the transition from observations from SOHO/MDI (prior to May 2010) and SDO/HMI (following May 2010).}
\label{fig:fX}
\end{figure}

The alternating toroidal field patterns clearly visible in the WSO observations are borne out also with considerably higher spatial resolution observations from space with SOHO/MDI and SDO/HMI shown in Fig.~\ref{fig:fX}, and Fig.~2 of \cite{2022ApJ...927L...2L} which, unlike our previous plots, are current to time of publication. In tandem, the three magnetograph observations illustrate the clear pattern of the ESC that is consistent with previous studies. Further, as we have discussed immediately above, we observe that another zero crossing of the toroidal magentic field at the equator, characteristic of a Hale Cycle terminator event, occurred very recently. In a forthcoming publication we will explore this event in detail (McIntosh et al. \-- in preparation).

\section*{Discussion}
A general criticism of the M2014 band-o-gram is that it was based on catalogued proxies of the photospheric magnetic field through chromospheric and coronal features. Those tracked features formed by the overlapping activity bands observed were not necessarily representative of the photospheric or interior magnetic field itself. It is clear from the WSO observations that, while comparison of the observed progression with the band-o-gram is still qualitative, that there is an overwhelming correspondence of the features observed in the WSO observations with those of the highly idealized band-o-gram. We note that a similar treatment of higher spatial resolution photospheric observations from the Mt Wilson Solar Observatory over a shorter timeframe yields similar correspondence \citep{2005ApJ...620L.123U}. 

\begin{figure}[ht]
\centering
\includegraphics[width=\linewidth]{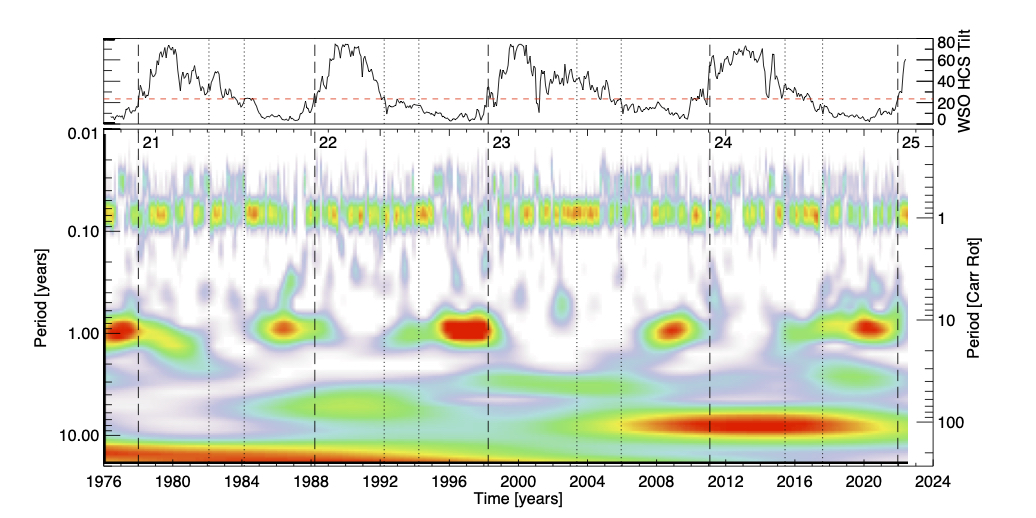}
\caption{(top) The tilt angle of the Heliospheric current sheet as measured by the WSO. (bottom) Morlet wavelet map of the interplanetary magnetic field polarity 1976--2022 showing periodicities from 4~days to 22~years. As in Figures~\ref{fig:f1} and~\ref{fig:f2}, the vertical dashed lines shown in each panel mark the times of the Hale Cycle termination events studied by M2019; the dotted vertical lines correspond to~0.4 and~0.6 of the cycles' duration from terminator to terminator \protect\citep[cf.][]{Leamon2022}. The dashed horizontal line in the HCS tilt panel is drawn at 23.4\degree{}; exceeding this value is a reasonable scalar proxy for the Hale Cycle terminator.}
\label{fig:fb}
\end{figure}

Further, it is known that the heliosphere exhibits a `sector' structure. The sector, or Hale sector, structure reflects the polarity of the heliospheric magnetic field relative to the solar direction in a state of either being ``away'' from or ``towards'' the Sun and expresses the largest spatial scales of solar magnetism and connectivity \citep[e.g.,][]{Hudson2014}. Since the earliest articles about sector structure \citep[e.g.,][]{1969JGR....74.5611R} the solar cycle has been noted to have a strong annual modulation around solar minimum. At that time the heliospheric current sheet (HCS) is so flat that for six months of the year (early December to early June) the Earth is at southern heliographic latitudes and the dominant polarity corresponds to the Sun's southern hemisphere. For the other six months of the year in these epochs the Earth almost exclusively samples the dominant polarity of the north, holding at a level of $\sim$85\%{} \citep[e.g.,][]{1975SoPh...41..461S}. The top panel of Fig.~\ref{fig:fb} shows the tilt of the HCS as computed by the WSO from 1976 to the present. The slowly evolving solar minimum behavior of the HCS is shown graphically in the lower panel of the figure \-- an adaptation of Fig~1 of \citet{2004GeoRL..3112808E}. The wavelet transform is used to illustrate the prominent periodicities in the sector structure \-- at approximately one Carrington rotation (CR) timescale, and the other at approximately one year. There are two clear results shown in Figure~\ref{fig:fb}: (1) the strongest signal at one year indeed corresponds to the times of extreme HCS flatness, but also that the strongest signal reverts to CR timescales at Hale Cycle terminators, when the tilt rises sharply with new and stronger new-cycle active regions emerging at mid latitudes. (2) the {\em onset\/} of the annual periodicity signal is at apprximately 0.4~cycles (first dotted vertical line) for even numbered cycles and at 0.6~cycles (second dotted vertical line) for odd numbered cycles. We reserve discussion of the the 22-year difference between odd and even numbered cycles to a manuscript in preparation, that looks at a longer epoch than that covered by the WSO we focus on. Nevertheless, this highly ordered large-scale sector structure is one more piece of evidence consistent with the data-inspired ESC schematic based on the timing of the Hale Cycle terminators.

\section*{Conclusion}
The meticulous daily synoptic scale observations of the WSO have captured two complete 22-year Hale cycles. These observations have permitted a mapping of the Sun's photospheric toroidal magnetic field component over that timeframe. Key features of the WSO observations compare directly to the data-inspired schematic of the ESC that was conceived to illustrate how the activity bands of the ESC can interact to shape the latitudinal progression of sunspot cycles and their amplitude. The WSO observations should unambiguously unify the Hale magnetic cycle and the ESC as being, physically, one and the same and indistinguishable. These low spatial resolution ground-based observations are corroborated by higher resolution space-based magnetographic observations from SOHO and SDO where all three identify zero-crossing events we associate as Hale Cycle terminators. As \cite{2010ASPC..428..109L} and M2014 inferred, there is predictive capability in these synoptic analyses through the ESC \--providing strong indicators of the current progression and potential evolution of upcoming solar activity at the decadal scale, beyond those amenable through the analysis of sunspots. This result demonstrates the intrinsic power of synoptic observations at a time when it is becoming increasingly difficult to sustain such efforts. 

\section*{Acknowledgements}
SMC is supported by the National Center for Atmospheric Research, which is a major facility sponsored by the National Science Foundation under Cooperative Agreement No. 1852977. RJL acknowledges support from NASA's Living With a Star Program. SMC and RJL acknowledge the grant of Indo-US Virtual Networked Center (IUSSTF-JC-011-2016) to support the joint research on ESCs. Stanford University operates WSO with funding provided by the National Science Foundation with Grant \#1836370. NASA funding for WSO ended in 2018. Historically WSO has been supported by NASA Heliophysics, the NSF, and the Office of Naval Research. Dr. J. Todd Hoeksema serves as WSO Director and we acknowledge his steadfast stewardship of the data and calibration. Sunspot data from the NOAA Space Weather Prediction Center and the World Data Center SILSO, Royal Observatory of Belgium, Brussels.

\section*{Conflict of Interest Statement}
The authors declare that the research was conducted in the absence of any commercial or financial relationships that could be construed as a potential conflict of interest.

\section*{Author Contribution Statement}
All authors conceived the experiment, P.S., L.S. and S.M. analyzed the results. 
S.M. created Figures~\ref{fig:f0}--\ref{fig:f2}; R.L. created Figure~\ref{fig:fb}.
All authors reviewed the manuscript.

\bibliographystyle{frontiersinSCNS_ENG_HUMS}


\end{document}